# Requirements Engineering Challenges in Building AI-Based Complex Systems

Hrvoje Belani
IT Division
Ministry of Health
Zagreb, Croatia
hrvoje.belani@miz.hr

Marin Vuković, Željka Car
Faculty of Electrical Engineering and Computing
University of Zagreb
Zagreb, Croatia
{marin.vukovic; zeljka.car}@fer.hr

*Abstract*—This paper identifies and tackles the challenges of the requirements engineering discipline when applied to development of AI-based complex systems. Due to their complex behaviour, there is an immanent need for a tailored development process for such systems. However, there is still no widely used and specifically tailored process in place to effectively and efficiently deal with requirements suitable for specifying a software solution that uses machine learning. By analysing the related work from software engineering and artificial intelligence fields, potential contributions have been recognized from agent-based software engineering and goal-oriented requirements engineering research, as well as examples from large product development companies. The challenges have been discussed, with proposals given how and when to tackle them. RE4AI taxonomy has also been outlined, to inform the tailoring of development process.

## I. Introduction

The term "requirements engineering" (RE) is widely used in the software engineering (SE) field to denote the systematic handling of requirements, which express the needs and constraints placed on a software product that contribute to the solution of some real-world problem [1]. RE is concerned with the elicitation, analysis, specification, and validation of software requirements as well as the management and documentation of requirements throughout software product life cycle (SPLC).

Artificial intelligence (AI) based on machine learning (ML), and in particular deep learning (DL), is today the fastest growing trend in software development, and literally in all areas of the modern society. However, a wide use of AI in many systems, complex and dependable systems in particular, is still not the case. These are the systems with simple components but complex overall behaviour due to the dependencies, competitions, relationships, or other types of interactions between the components or between a given system and its environment [2]. Along with a shortage of expertise, the challenges for managing AI-based complex and dependable systems are enormous, though less known, and in general underestimated [3].

The scope of this paper is to target challenges that arise in RE as a disciplined approach to develop AI-based complex systems. Related work dealing with AI development challenges and approaches has been informally looked into, as well as some RE approaches that could serve as candidates to align with AI development efforts.

Based on the discussion, the RE4AI ("RE for AI") taxonomy has been outlined, considering to be useful to inform the tailoring of AI development process. The elements of the taxonomy have been recognized from the AI field – data, model and system, and aligned with the typical RE activities. The purpose of the proposed taxonomy is to help practitioners and researchers to focus on certain challenges upfront and tailor their AI research approaches and development processes accordingly.

This work has been partly motivated by [3] as well as the first author's participation in [4]. Method and settings are presented in section II, while section III provides an overview of related work, with discussion on challenges and approaches given in section IV. Section V outlines the RE4AI taxonomy, and section V concludes the paper.

## II. Method and Settings

The purpose of every software development process, set in the given organizational environment and under the dynamic circumstances, should be to propose the right level of discipline, drive knowledge and enhance communication while not repressing innovation and creativity of the people playing the process roles. In that sense, software process tailoring to suit its context is recognized to be a key mechanism to address the challenges faced by a project, by identifying a set of environmental factors, challenges, project goals, process-tailoring strategies, and their influences on each other [5]. The duality of the software process has been recognized, showing how the context and the tailoring decisions dynamically interact with each other and construct the context in which the project is developed and the process is tailored. Even more, when dealing with complex software systems development and integration, there is a reported need for a hybrid process model for project management and software development in agile environment, which is in a certain manner suitable for agile conditions and adaptive behaviour [6].

The need for RE became obvious as the systems engineering discipline developed, but with emerging new technologies as well as business needs, new approaches were needed [7]. As one of the approaches to effectively manage the requirements for complex systems, the model-driven requirements engineering (MDRE) has been established, covering both business modelling and analysis modelling activities [8]. From the year 2000 onwards it has been found that the AI area is in the sustainable development and its impact continues to grow [9].

## III. RELATED WORK

Building AI-based complex systems surely goes beyond using specific AI algorithms, and the development itself is becoming more complex since data needed and algorithms implemented become dependent. The system generally consists of a variety of subsystems, some of which are data-centric while others will be model-driven. Part of the challenge of combining such subsystems, as reported in [10], is non-existence of widely used software development paradigm that deals explicitly with autonomous systems. Nevertheless, there is an agent-based software engineering (ABSE), being somehow present in AI research for more than two decades, which can provide a solid starting point for reasoning about the system as a whole and about the interactions between the various subsystems. ABSE offers a variety of architectural approaches, both for individual agents and systems with multiple agents interacting. An intelligent autonomous agent is considered to be a subsystem able to make an acceptable decision about what action to perform next to its environment, in time for this decision to be useful [11].

Grouping the agents based on their degree of perceived intelligence and capability results with the following five classes [12]: simple reflex agent, model-based reflex agent, goal-based agent, utility-based agent and learning agent. Interesting here is a goal-based agent, being a model-based agent upgraded by using goal information describing desirable situations and selecting from a set of possible actions the one that improves the progress towards the goal, not necessary the best one.

Except from providing the potential contributions from ABSE to requirements for AI, in terms of defining the properties AI modules or subsystems should have, it is crucial to enhance the organization's business intelligence and analytics capabilities to support ML, as reported in [13], by providing the following: (1) updating the data organization layer in end-to-end analytics architectures to support data preparation for ML algorithms; (2) incorporating a SPLC that supports learning models when the organization plans to aggressively build custom ML algorithms and applications; (3) choosing an ML platform that supports and interoperates with multiple ML frameworks when the organization plans to leverage service providers or commercial off-the-shelf solutions; (4) focusing on storage and compute clusters to support ML capabilities (e.g. choosing the public cloud over own IT infrastructure).

### A. AI Development Challenges at Large

One of large organizations with significant experience in AI development, as reported in [14], conducted a two-phase study with a set of interviews to gather the major topics and a wide-scale survey about the identified topics, in order to observe their software teams as they develop AI-based applications. They found that various teams have united the new workflow into pre-existing, well-evolved, agile-like SE processes, providing insights about several engineering challenges that organizations may face in building AI-based complex systems [14]: end-to-end pipeline support; data availability, collection, cleaning, and management; education and training; model debugging and interpretability; model evolution, evaluation, and deployment; compliance; varied perceptions.

### B. Optimizing and Combining AI Development Approaches

Going more precisely to the first phases of SPLC, the multi-case study research in two industrial domains – embedded systems and online companies [15], identifies three distinct approaches to AI-based software development, along with the given typical problems that organizations experience when using the wrong approach for the wrong purpose:

1) *Requirement driven development:* software is built to specification and product management is responsible for collecting and specifying requirements as input for the development teams,
2) *Outcome/data driven development:* development teams receive a quantitative target, i.e. an outcome, to realize and are asked to experiment with different solutions to improve the metric,
3) *AI driven development:* the organization has a large data set available and uses AI techniques such as ML and DL to create components that act based on input data and that learn from previous actions.

Based on the conducted research, holistic development framework has been proposed in [15], outlining the identified purposes for which each of the development approaches is well suited for: (1) regulatory features, for competitor parity features and for commodity features, (2) value hypotheses, development of new "flow" features, i.e. features used frequently by customers and for innovation, and (3) when aiming to minimize prediction errors, when there are many data points and when there is a combinatorial explosion of alternatives.

### C. Challenges in SE for DL

Investigating more into software engineering challenges of deep learning, the interpretive research approach in close collaboration with companies of varying size and type, has been reported in [16], and used seven ML projects to identify twelve main challenges categorized into the three areas:

- *Development challenges:* experiment management, limited transparency, troubleshooting, resource limitations, testing,
- *Production challenges:* dependency management, monitoring and logging, unintended feedback loops, glue code and supporting systems,
- *Organizational challenges:* effort estimation, privacy and data safety, cultural differences.

## IV. DISCUSSION ON CHALLENGES

The challenges on AI-based development identify them throughout the SPLC, not only at the earliest phases dealing with requirements. This seems logical because introducing ML subsystem into the complex system demands interventions to the SE processes on many levels, especially when dealing with datasets availability, ML models versioning and the whole system performance, including dependence on the hardware. Still, introducing the ML subsystem as a "black-box" element into the SE processes seems to violate the traceability property of system requirements, concerned with recovering the source of requirements and predicting the effects of requirements, which is fundamental for impact analysis when requirements change.

A software requirement should be traceable backward to the system requirements that it helps satisfy, as well as to stakeholders that motivated it. Also, is should be traceable forward into design entities – code modules that implement it or the test cases that verifies it. These and other trade-offs have been reported [17] using the framework of technical debt, and highlighting several ML specific risk factors and design patterns to be avoided or refactored where possible. These include, among others:

- *Entanglement:* ML models create entanglement and make the isolation of improvements effectively impossible. The CACE (Changing Anything Changes Everything) principle is in force,
- *Undeclared consumers:* If a prediction from a ML model is made accessible to a wide variety of systems, either at runtime or by writing to logs, that may later be consumed by other systems, called undeclared consumers, and served as an input to another component of the system,
- *Data dependencies:* They can be: (1) unstable, meaning that they qualitatively change behaviour over time; (2) underutilized, including input features or signals that provide little incremental value in terms of accuracy. It is costly and can creep into a ML model like e.g. legacy (early in development) features or bundled (hidden) features; (3) difficult of performing static analysis and track the use of data in a system; (4) causing correction cascades, caused by learning a model with a solution to a similar problem, making a correction model. However, future improvements (cascades) of that model are significantly more expensive to analyse,
- *System-level design anti-patterns:* glue code, pipeline jungles, dead experimental code paths, and configuration debt.

## V. Towards RE4AI Taxonomy

Looking for adequate research already done in the RE field that could possibly serve as a starting point to tackle all these challenges, the area of goal-oriented RE (GORE) has been identified [18], as goal modelling has been adapted and applied to many sub-topics within RE and beyond, such as agent orientation, aspect orientation, business intelligence, model-driven development, and security. Although goals have long been used in AI [19] and then modelled explicitly in requirements models to provide a criterion for requirements completeness [20], direct mapping to software prototypes with an agent-oriented architecture which can be executed for requirements validation and refinement [21], or even model-based framework for designing self-adaptive software systems that can guarantee a certain level of requirements satisfaction over time by dynamically composing adaptation strategies when necessary [22], it seems there is no research on goal modelling, taking a model of AI-based complex system that views ML sub-systems as goal-seeking entities, extensively used by AI development practitioners. The lack of specific expertise among AI and SE practitioners could be relaxed by offering a taxonomy.

Taking the analysed related work and discussed challenges into account, the RE4AI taxonomy has been outlined, providing summarized view on challenges posed to RE towards building AI-based complex systems. A broader taxonomy of SE challenges for ML systems exists [23], but the following one, as shown in Table 1, focuses on roles, processes and artefacts regarding the RE activities that need to be conducted in order to implement AI-based system.

TABLE I. RE4AI Taxonomy Outline with Mapped Challenges

| RE4AI RE activities | AI-related entities | | |
|---|---|---|---|
| | *data* | *model* | *system* |
| elicitation | - availability of (large) datasets<br>- requirements analyst upgrade | - lack of domain knowledge<br>- undeclared consumers | - how to define problem/scope<br>- regulation (e.g. ethics) not clear |
| analysis | - imbalanced datasets, silos<br>- role: data scientist needed | - no trivial workflows<br>- automation tools needed | - no integration of end results<br>- role: business analyst upgrade |
| specification | - data labelling is costly, needed<br>- role: data engineer needed | - no end-to-end pipeline support<br>- minimum viable model useful | - avoid design anti-patterns<br>- cognitive / sys. architect needed |
| validation | - training data critical analysis<br>- various data dependencies | - entanglement, CACE problem<br>- high scalability issues for ML | - debugging, interpretability<br>- hidden feedback loops |
| management | - experiment management<br>- no GORE-like method polished | - difficult to log&reproduce<br>- DevOps role for AI needed | - IT resources limitation, costs<br>- measuring performance |
| documentation | - data & models visualisation<br>- role: research scientist useful | - datasets and model versions<br>- education and training of staff | - feedback from end-users<br>- development method (agile?) |
| all of the above | - dataset privacy and data safety; - data dependencies | | |

Each of the challenges has been mapped to the certain AI-related entity, as well as given RE activity, in order to inform organizing and tailoring of the AI development process, focusing especially on requirements-related efforts. For example, execution of hidden feedback loops represents a challenge when AI-based complex system goes into production and should be addressed more from the requirements validation perspective, in order to ensure they are minimized.

Along with the challenges given in the RE4AI taxonomy outline, there is a number of gaps recognized from development for AI practice that need to be addressed, between:

- Software engineers and data scientists (*the skill gap*),
- Available and desirable (big) datasets (*the data gap*),
- Prototyping and full lifecycle support (*the engineering gap*).

## VI. Conclusions

The most attention of the AI industry is currently given to ML as a data-driven approach, due to the information technology (IT) infrastructural developments offered today, like fast processing power and non-expensive data storage. One of the areas most attracted to AI application is healthcare [24], yet with notable obstacles, e.g. lack of mandatory standards or continuous data exchange.

This paper represents a contribution to the topic of RE discipline for building AI-based complex systems, outlining RE4AI taxonomy. GORE is potentially applicable approach to build upon, but further analysis should look deeper into the applicability of existing GORE frameworks and methods [19]. A new system design paradigm [25], which combines data-driven and model-based design (DMD), should be looked into more details, as well as its fitness with RE4AI taxonomy.

If applied to healthcare domain, the RE4AI taxonomy may be helpful in tackling some of the emerged challenges [26], such as distributional shift and unsafe failure mode. The first one deals with imbalanced datasets and needs to be addressed early in RE phases, and the second should be concerned through all activities as it impacts the whole system behaviour.

ACKNOWLEDGMENT

This work has been carried out during HelloAI Summer School, which has been funded by EIT Health, and organized by General Electric (GE) Healthcare (Budapest, Hungary), KTH Royal Institute of Technology (Stockholm, Sweden), and LEITAT (Barcelona, Spain). EIT Health is supported by European Institute for Innovation & Technology (EIT), a body of the European Union (EU). The authors would like to thank Dr. Peter Bencsik, MD, and Krešimir Friganović, MSc EE IT, for their cooperation and support.